\documentstyle[12pt,epsfig,aaspp4,flushrt]{article}

\begin{document}

\def\lsim{\mbox{${\scriptstyle \buildrel < \over \sim}$}}
\def\gsim{\mbox{${\scriptstyle \buildrel > \over \sim}$}}
\def\HST{{\it Hubble Space Telescope}}


\title{
The Redshift Distribution and Luminosity Functions \\
of Galaxies in the Hubble Deep Field \\ }

\author{Stephen D. J. Gwyn and F. D. A. Hartwick}
\affil{Department of Physics and Astronomy, University of Victoria,\\
Box 3055, Victoria, British Columbia, V8W 3P6, Canada \\
gwyn, hartwick@uvastro.phys.uvic.ca\\
}

\begin{abstract}

Photometric redshifts have been
determined for the galaxies in the
Hubble Deep Field.
The resulting redshift distribution
shows two peaks:
one at $z\sim0.6$ and one at $z\sim2.2$.
Luminosity functions derived from
the redshifts show strong luminosity
evolution as a function of redshift.
This evolution is
consistent with the Babul \& Rees (1992)
\nocite{br92}
scenario wherein
massive galaxies form stars at 
high redshift while star formation
in dwarf galaxies is delayed until
after $z=1$.
\end{abstract}

\section{Introduction}
\label{sec:intro}

The Hubble Deep Field 
(HDF) optical images are
the deepest yet obtained.
Objects as faint as $I_{ST}$=28.5
\footnote{for simplicity, $U_{ST}$,$B_{ST}$,$R_{ST}$ and $I_{ST}$ 
will be used to denote magnitudes in the F300W, F450W, F606W and F814W
bands respectively. 
The ST zero-point system is used unless otherwise
specified.}
can be detected 
at the $10\sigma$ level.
At this point in time, 
only a few spectroscopic redshifts have been measured
for the brighter galaxies
and none for the
faintest galaxies in these images. 

Photometric redshifts 
(see, for example, Gwyn 1995; Connolly et al. 1995; Koo 1985)
\nocite{gwyn95,evil,pmzm} 
can be measured much faster and to much fainter
magnitudes than
their spectroscopic counterparts.
Because the wavelength bin-size
in photometry is generally so much larger
than in conventional spectroscopy ($\sim$1000\AA~
{\em vs.} 1--2\AA),
far shorter exposure times are required 
to measure redshifts (but with a sacrifice in accuracy).

Photometric redshifts have been calculated for
the galaxies brighter than
$I_{ST}=28$ in the Hubble Deep Field.
This paper presents the redshift
distribution for this sample.
Also presented are
luminosity functions
as a function of redshift out to $z=5$.

\section{The photometric redshift technique}
\label {sec:phot}

The photometric redshift technique can be divided into three steps:
First, 
the photometric data for each galaxy in the fields are 
converted into spectral energy distributions (SED's).
Second, 
a set of template spectra of all Hubble types and redshifts 
ranging from $z=0$ to $z=5$ is compiled.
Third,
the spectral energy distribution derived from the observed 
magnitudes of each object is compared to each template 
spectrum in turn. 
The best matching spectrum, and hence the redshift, 
is determined by minimizing
$\chi^2$.
In the following subsections, each of these steps is examined.

\subsection{Photometry to spectral energy distributions}
\label{ssec:seds}

The magnitude in each bandpass is converted to a flux 
(power per unit bandwidth per unit aperture area) 
at the central or effective wavelength, $\lambda_{cen}$, 
of the bandpass.
When the flux is plotted against wavelength 
for each of the bandpasses, 
a low resolution spectral energy distribution is created.

The following equation converts magnitudes 
in some filter, $f$, to fluxes:
\begin{equation}
\label{flux2mag}
F_f=F_0 10^{-m_f/2.5},
\end{equation}
where 
$m_f$ is the apparent magnitude,
$F_f$ is the flux in units of W \AA$^{-1}$ m$^{-2}$ and
$F_0$ is the flux zero-point of that filter system in the same
units.
This equation is simple to use as long as
the flux zero-point is known.
Unlike almost all other magnitude systems ({\it e.g.}~the 
Johnson/Cousins
UBVRI system)
the flux zero-points for the ST magnitude system are the same
for all filters, making the conversion of magnitudes to fluxes
straightforward.

\subsection{The template spectra}
\label{ssec:temp}

The template spectra were produced from the
models of 
Bruzual \& Charlot (1993).
\nocite{bc93}
They showed that their models faithfully reproduce the spectra of local galaxies.
More recently, Steidel et al. (1996a)
\nocite{steidpp} found that
high redshift galaxies are also well described by the models.

The templates were produced in
a two-dimensional grid
with galaxy type (ranging from Sm to E/S0) in one dimension
and redshift (ranging from  $z=0$ to $z=5$) in the other.
The galaxy templates included evolution 
according to the Bruzual \& Charlot (1993) models.
\nocite{bc93}
It was assumed that they formed 0.5 Gyr
after the Big Bang and evolved
reaching an age of 15 Gyr at the present
epoch.
A cosmology where $H_0=50$ km sec$^{-1}$ Mpc$^{-1}$, $\Omega_0=0.3$, and $\lambda_0=0$ was used to convert redshift
to the evolutionary epoch of the models and in the later
calculation of the luminosity functions. 

To represent the different galaxy types
an interpolation was made between the
instantaneous burst model (for early-type galaxies) and the
constant star formation rate model (for late-type galaxies).
These interpolated spectra were then redshifted.
The redshifted spectra were reduced to the passband
averaged fluxes 
at the central wavelengths
of the passbands, in order
to compare the template spectra with the
SED's of the observed galaxies.

\subsection{Comparing the templates to the SED's}
\label{ssec:comp}

Given a spectral energy distribution of a galaxy of
unknown redshift and a set of templates, 
the next step is to compare the SED to each of the
templates in turn and to the template which
most closely matches the SED.
The degree to which each template matches the observed SED
is quantified in the following manner:
\begin{equation}
\label{eqn:chimin}
\chi^2=\sum_{i=1}^{N_f} {(F_i-\alpha T_i)^2 \over \Delta F_i^2},
\end{equation}
where $N_f$ is the number of filters in the set,
$F_i$ and $\Delta F_i$ are respectively the flux and the uncertainty in
the flux in each bandpass of the observed galaxy, 
$T_i$ is the flux in each bandpass of the template being considered,
and $\alpha$ is a normalization factor.
A normalization factor is necessary 
to compare properly the galaxies and the templates because
the fluxes of the galaxies are
very small relative to the fluxes
of the templates.
Each template is then compared to the target galaxy SED and
the smallest value of $\chi^2$ is found.
The best matching template gives 
$z_{phot}$, the sought-after photometric redshift of the galaxy.

\subsection{Accuracy}
\label{ssec:res}
The photometric redshift technique has been tested 
with simulations and photometric observations
of galaxies with known spectroscopic redshifts.

\subsubsection{Effects of photometric errors}
\label{sssec:phote}
The simulations take the following form:
A spectral energy distribution
is chosen at random from the template SED's available.
The SED is reduced to fluxes at the
central wavelength of the passbands
of the filter set to be tested.
To simulate observational errors,
a small random number with a Gaussian distribution 
is added to the fluxes in each band of the chosen template.
Photometric redshifts are determined from
the resulting simulated photometry
in the manner described above.
These photometric redshifts are compared
with the redshifts of the templates from
which they were derived.

Not surprisingly,
when the errors in the photometry are zero,
the photometric redshift technique
always picks the correct redshift.
As the errors in the photometry 
increase, so do the errors
in the photometric redshifts in the form
of the standard deviation of redshift residuals.
The simulations indicate that 
the uncertainties in the photometric redshifts, $\sigma_{pz}$,
increase with redshift.
For $z<1.5$ they are only $\sigma_{pz}\simeq0.07$;
for $1.5<z<5$ they are $\sigma_{pz}\simeq0.20$.

\subsubsection{Effects of variations in evolutionary history}
\label{sssec:eve}

As a check on the sensitivity of the derived
redshifts to a particular evolutionary model,
photometric redshifts were calculated for the HDF galaxies using
non-evolving templates, {\em i.e.} templates
derived by redshifting the SED's of local galaxies.
Although in general these photometric
redshifts were different from 
those derived from the evolving templates,
no systematic difference was found.
These differences can be attributed
to the different evolutionary histories
of the individual galaxies.
The differences represent uncertainties in the redshifts
of between $\sigma_{pz}\simeq0.05$ (low redshift) and 
$\sigma_{pz}\simeq0.2$ (high redshift).
The lack of a systematic difference can be understood
if a portion
of the evolution in the SED's takes the
form of a change in galaxy spectral type.
Further, the photometric redshift technique
is sensitive to the location of large breaks
(at 4000\AA~ and 912\AA) which change with
redshift, not type or evolutionary state. 

\subsubsection{Effects of internal absorption by dust}
\label{sssec:duste}

The reddening effects of dust are not
accounted for in the Bruzual \& Charlot (1993) models.
\nocite{bc93}
In order to evaluate the effects of dust on the
derived redshifts, moderate amounts of internal reddening were added
to the template SED's.
Random differences of $\sigma_{pz}\simeq0.05$ (independent of redshift)
were found between the redshift derived from these templates
and those derived from un-reddened templates,
but again no systematic differences were evident.

\subsubsection{Comparison with available spectroscopic redshifts}
\label{sssec:compe}

Some spectroscopic redshifts are
available for the HDF:
those of Cowie (1996) (23 galaxies with $z<1.5$ at this point in time)
\nocite{cow96}
and Steidel et al. (1996b) (5 galaxies with $z>1.5$).
\nocite{steidsp}
A comparison of these spectroscopic
redshifts and photometric redshifts
shows that
the uncertainties in the photometric
redshifts at high redshift ($z>1.5$) are
$\sigma_{pz}\simeq0.5$. 
At low redshift, 
two out of 23 galaxies have 
redshift discrepancies greater than $\Delta z=0.5$
while the remaining 21 have a standard deviation of $\sigma_{pz}\simeq0.2$.

When one adds the above uncertainties due to photometry,
evolution and moderate internal reddening one finds that
$\sigma_{cal}\simeq0.1$ ($z<1.5$) and
$\sigma_{cal}\simeq0.3$ ($z>1.5$).
The observed uncertainties 
are clearly much larger:
$\sigma_{obs}\simeq0.2$ ($z<1.5$) and
$\sigma_{obs}\simeq0.5$ ($z>1.5$).
However, it is not difficult to imagine additional sources of
uncertainty. Among them are absorption by the intervening IGM
(both gas and dust) and evolutionary metal abundance effects.
Both of these effects are difficult to quantify and 
given the primarily statistical nature of this
investigation we rely on the
empirically derived uncertainties.

\section {The redshift distribution and luminosity function}
\label {sec:lumf}
The positions of the galaxies in the HDF
images were taken from the catalogs
of Couch (1996\nocite{couch96}).
Magnitudes were measured for all galaxies
from the HDF Version 2 images
through a 0.2 arc-second radius aperture
using IRAF.
The photometric redshift technique was used on all galaxies
brighter than $I_{ST}>28$. 
A small fraction (6\%) of the
galaxies have no $U_{ST}$ magnitudes;
redshift for these galaxies were calculated
using only the $B_{ST}R_{ST}I_{ST}$ photometry.
These redshifts are somewhat less accurate
but have the same distribution
as the 4-colour redshifts.
Three galaxies have no measurable $U_{ST}$ or $B_{ST}$ flux.
Photometric redshifts cannot be calculated for them;
they were ignored in the following discussion.

The redshift distribution is shown in Figure~1.
The distribution shows two peaks, one at a redshift
of about $z\simeq0.6$ and another at $z\simeq2.2$.
The redshift distribution of the 
of the galaxies without $U_{ST}$ magnitudes
is shown as a dashed line.
Compared to spectroscopic redshifts, where typically 
$\Delta z \sim 0.001$, the uncertainties associated 
with photometric redshifts are large.
Note, however, that the width of the bins in this
histogram (0.4 in $z$) are comparable to the errors
expected in the photometric redshifts 
($\sigma_{pz}\sim0.2-0.5$).
Simulations indicate that this level of 
precision is sufficient to calculate accurate 
luminosity functions.
Indeed, photometric redshifts have been used at 
$z\sim0.35$
to calculate LF's identical to those determined using 
spectroscopic redshifts (Gwyn 1995\nocite{gwyn95}).

The luminosity functions (LF's) were constructed 
using the $1/V_a$ method. 
The $1/V_a$ method is well known 
and has been described in detail elsewhere 
(\nocite{schmidt68,cfrs6}Schmidt 1968; Lilly et al. 1995b) 
so the following description is brief.

$V_a$ is defined as the volume accessible to a 
galaxy given its absolute magnitude and the limits 
defining the sample in which it is found.
Formally, 
\begin{equation}
V_a=\int_{z_{min}}^{z_{max}} {dV \over dz} dz,
\end{equation}
where $dV/dz$ is the co-moving differential volume element.
The limits, $z_{min}$ and $z_{max}$ can be fixed 
(as in a volume limited sample), 
but for a magnitude limited sample they must be 
determined for every galaxy. 
Given the absolute magnitude, $M$, of each galaxy 
and the limiting apparent magnitude of the sample, 
$m_{lim}$, the limiting redshift at which the galaxy 
would still be in the sample, $z_{lim}$, can be determined.
The luminosity function, $\Phi$, is the sum of the 
inverse of the
accessible volumes ($\sum 1/V_a$) normalized to 
the angular
area surveyed.

The luminosity function was calculated for three redshift regions:
$z<1$, $1<z<3$ and $3<z<5$.
In each case, $z_{min}$ was the lower bound of each redshift region 
($z=$0, 1 and 3 respectively).
The minimum of the upper bound of each redshift region 
($z=$1, 3 and 5 respectively)
or $z_{lim}$ 
(as determined for each galaxy using $I_{ST}=28.0$
for $m_{lim}$)
was used for $z_{max}$.

The luminosity functions thus calculated are shown in Figure~2.
In order to compare the
luminosity functions with local LF's, they are shown
in the (Johnson) $B$ band.
Note that it is not necessary when using the $1/V_a$ method
to use the same filter to calculate both the $V_a$'s 
and the absolute magnitudes.

Shown for comparison as a solid line 
is the local luminosity function of Loveday (1992\nocite{love92}).
The lowest redshift division ($z<1$, dotted line) shows the 
steep faint
end of the local luminosity function.
The next highest redshift division ($1<z<3$, dashed line) lacks
this faint tail; the space density of faint galaxies
seems depressed with respect to the $z<1$ LF.
The bright end is about 4 magnitudes brighter than
the local luminosity function.
The bright end is fainter by about 0.5 magnitudes in the highest
redshift interval ($3<z<5$, dot-dashed line); the faint end
drops off at a brighter magnitude.
Although different cosmologies change the positions
(and, to a lesser degree, the shape) of the luminosity functions,
changing the parameters does not greatly alter the relative positions
of the LF's at various redshifts. 

These variations in the luminosity function with redshift
can be explained if there are two epochs at which
galaxies undergo their first major burst of star formation.
The first occurs before $z \gsim 2.2$; at this time the larger 
galaxies start forming stars. 
The galaxies do not all burst simultaneously.
The stars of some galaxies first turn
on at $z\sim5$, some 
form as late as $z\sim2$ while most form at 
$z\sim2.2$, corresponding to the peak of the
redshift distribution shown in Figure~1.
Depending on the fraction of gas that is
converted into stars in the initial starburst,
a fading of 4 magnitudes since $z\sim2.2$
is entirely consistent with the Bruzual \& Charlot
(1993\nocite{bc93}) models.

The heuristic model of
Broadhurst, Ellis \& Shanks (1988)
\nocite{bes88}
of starbursting dwarfs
at moderate redshifts
was put on
a more physical basis
by Babul \& Rees (1992) and Babul \& Ferguson (1996)
\nocite{br92,bf96}
who postulated that
star formation in dwarf galaxies
is delayed until $z\sim1$ by
photoionization of their gas by inter-galactic
ultraviolet radiation.
This second epoch of star formation
explains the observed excess at moderate redshifts
of faint blue galaxies (termed ``boojums'' for
``blue objects observed just undergoing moderate
starburst'' by Babul \& Rees, 1992).
These galaxies end up on the steeply rising
faint tail of the lowest redshift luminosity function.
This scenario 
would also explain why the higher redshift
luminosity functions in Figure~2
have depressed faint ends:
the galaxies that would populate the faint end
haven't turned on yet.

\section{Summary}
\label {sec:conc}
Using photometric redshifts,
a redshift distribution has been determined
for the Hubble Deep Field. 
It shows two peaks: 
one at $z\simeq0.6$ and another at $z\simeq2.2$.
Luminosity functions have been calculated using
these redshifts.
The LF's show strong evolution:
the brightest galaxies are 4 magnitudes 
brighter than their present day counterparts 
and the faint galaxies are fewer in number.
The double-peaked redshift distribution
and the evolution of the luminosity function
can be understood if larger galaxies
form stars early at $z\sim3$ and
if star formation is delayed in the dwarf
galaxies until after $z\simeq1$.

\acknowledgements
Thanks are due to D. Crampton for suggesting
the application of our photometric redshift technique
to the Hubble Deep Field.
F.D.A.H. gratefully acknowledges financial support
for this project through an NSERC operating grant.

{\raggedright

}

\newpage



\newpage
\begin{figure}[ht]
\plotone{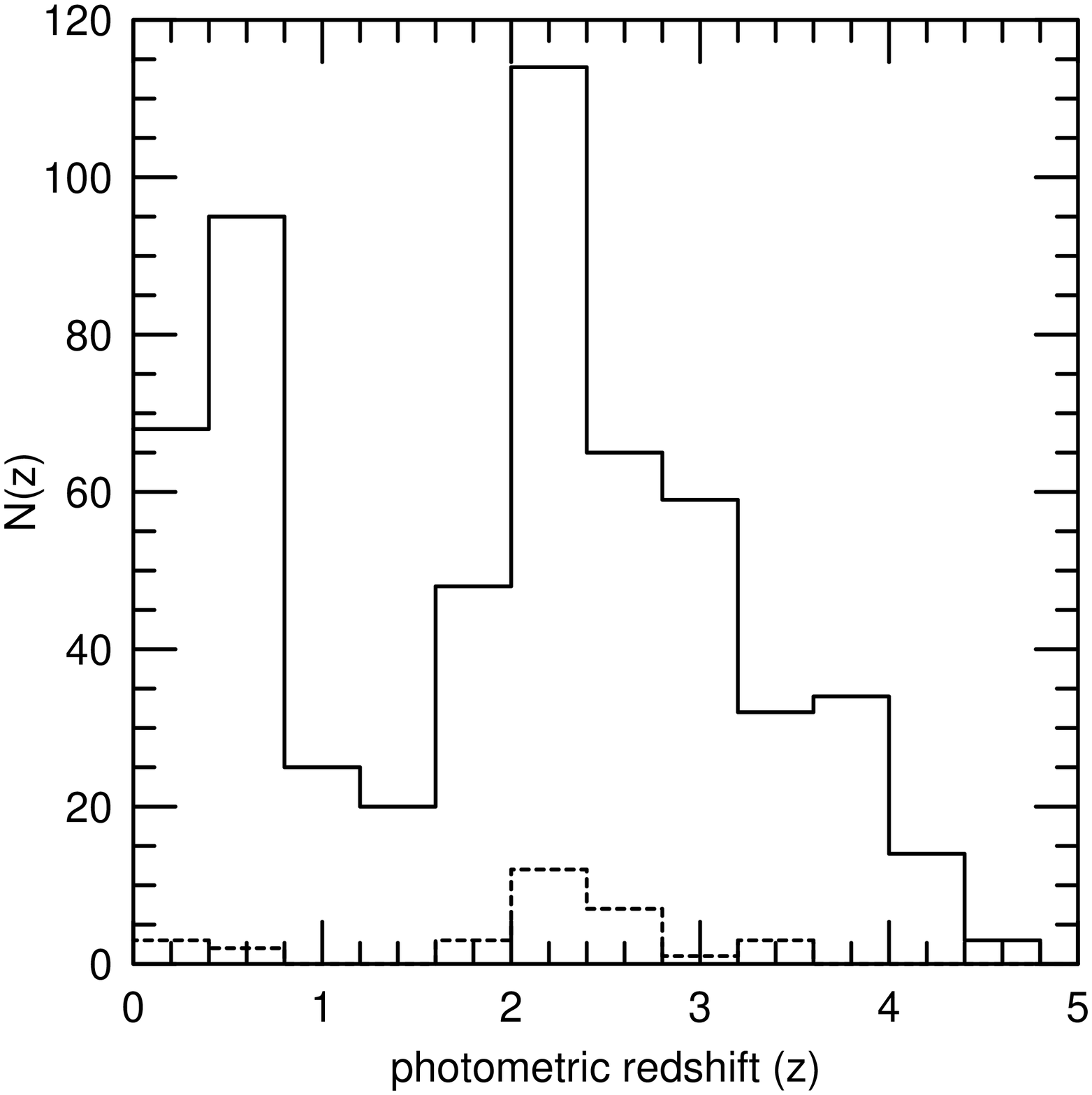}
\caption{
The HDF redshift distribution.
Two peaks are visible in the distribution:
one at $z\sim0.6$ and one at $z\sim2.2$.
The redshift distribution of those galaxies
for which $U_{ST}$ is not available
is shown as a dashed line.
}
\end{figure}

\begin{figure}[ht]
\plotone{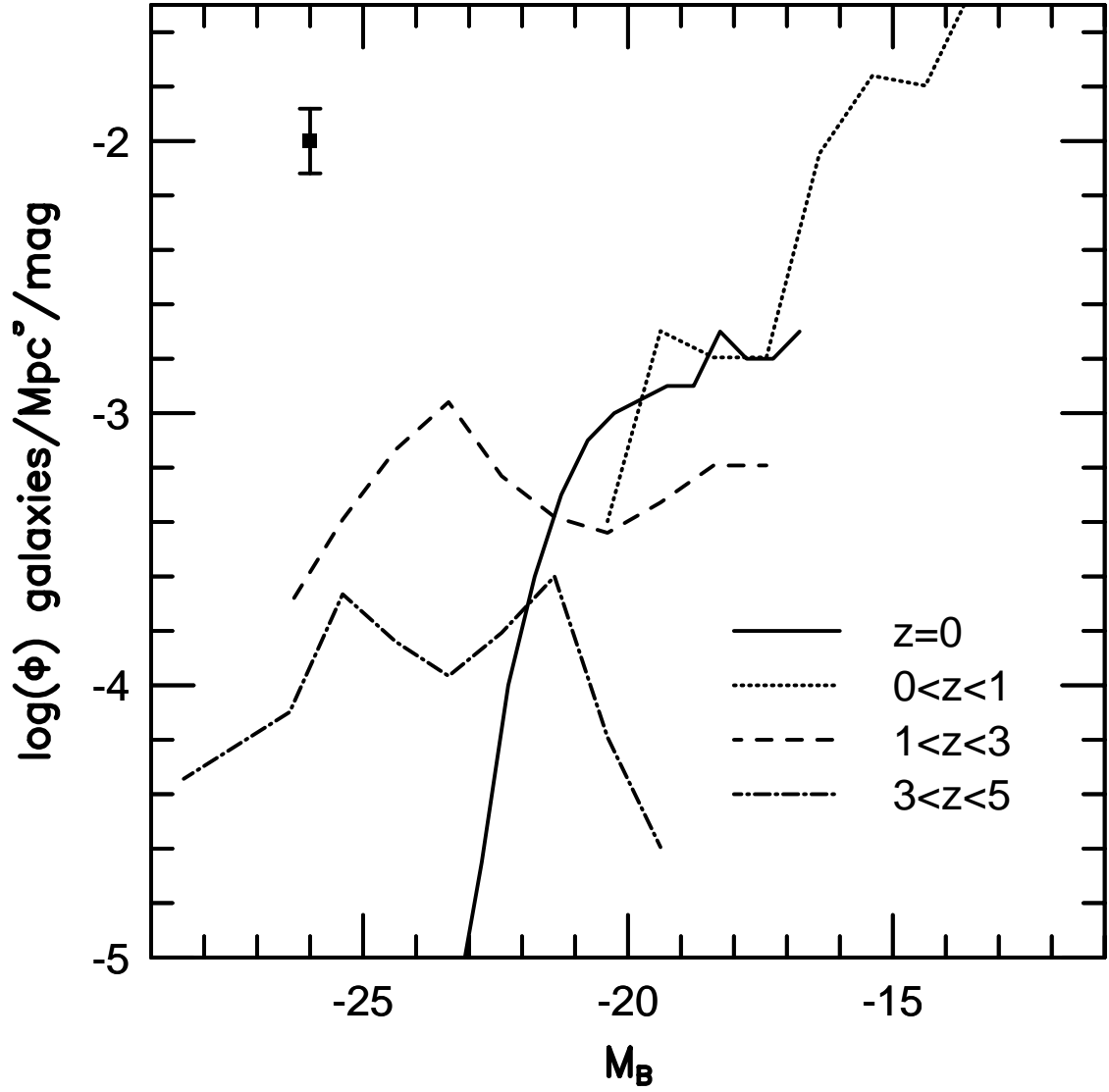}
\caption{
The evolution of the (Johnson) $B$ luminosity function
with redshift.
The local LF is that of Loveday et al. (1992).
The error bar shows the typical 
$1/ \sqrt {N} $
$N^{-{1 \over 2}}$
uncertainties which represent a lower limit to the true uncertainties.
}
\end{figure}


\begin{thebibliography}{}

\bibitem[Babul \& Ferguson, 1996]{bf96}
Babul, A. \& Ferguson, H.~C. 1996,
   \apj,  458, 100

\bibitem[Babul \& Rees, 1992]{br92}
Babul, A. \& Rees, M.~J. 1992,
   \mnras,  255, 364

\bibitem[1988]{bes88}
Broadhurst, T.~J., Ellis, R. S., \& Shanks, T. 1988, 
   \mnras, 235, 827

\bibitem[1993]{bc93}
Bruzual, G.~A. \& Charlot, S. 1993,
   \apj,  405, 538

\bibitem[(Connolly et~al., 1995)]{evil}
Connolly, A.~J., Csabai, I., Szalay, A.~S., 
Koo, D.~C., Kron, R.~G., \& Munn,
  J.~A. 1995,
   \aj,  110, 2655

\bibitem[1996]{couch96}
Couch, W.~J. 1996,
 Public communication,
 [http //ecf.hq.eso.org/hdf/catalogs]

\bibitem[1996]{cow96}
Cowie, L.~L. 1996,
 Public communication,
 [http://www.ifa.hawaii.edu/~cowie/hdf.html]

\bibitem[(Gwyn, 1995)]{gwyn95}
Gwyn, S. D.~J. 1995, unpublished M.Sc. thesis, University of Victoria

\bibitem[(Koo 1985)]{pmzm}
Koo, D.~C. 1985,
   \aj,  90, 418

\bibitem[Lilly et~al., 1995b]{cfrs6}
Lilly, S.~J., Tresse, L., Hammer, F., Crampton, D., \& Le F\`evre, O. 1995b,
   \apj,  455, 108

\bibitem[1992]{love92}
Loveday, J., Peterson, B.~A., Efstathiou, G., \& Maddox, S.~J. 1992,
   \apj,  390, 338

\bibitem[Schmidt, 1968]{schmidt68}
Schmidt, M. 1968,
   \apj,  151, 393

\bibitem[1996a]{steidpp}
Steidel, C.~C., Giavalisco, M., Pettini, M., Dickinson, M., \& Adelberger,
  K.~L. 1996a,
 Preprint [astro-ph/9602024]

\bibitem[1996b]{steidsp}
Steidel, C.~C., Giavalisco, M., Dickinson, M., \& Adelberger,
  K.~L. 1996b,
 Preprint [astro-ph/9604140]


\end{thebibliography}
\end{document}